\documentstyle[preprint,prb,eqsecnum,aps]{revtex}
\begin{document}
\draft
\title{Adsorption of $^4$He on a single C$_{60}$}
\author{Leszek Szybisz$^{1,2,}$\cite{szybco} and Ignacio
Urrutia$^{1,}$\cite{urrucic}}
\address{$^1$Laboratorio TANDAR, Departamento de F\'{\i}sica,
Comisi\'on Nacional de Energ\'{\i}a At\'omica,\\
Av. del Libertador 8250, RA-1429 Buenos Aires, Argentina}
\address{$^2$Departamento de F\'{\i}sica, Facultad de
Ciencias Exactas y Naturales,\\
Universidad de Buenos Aires,
Ciudad Universitaria, RA-1428 Buenos Aires, Argentina}
\date{\today}
\maketitle
\begin{abstract}
The adsorption of $^4$He inside and outside a single fullerene
C$_{60}$ is studied. A physisorption potential is proposed. The
energetics and structural features of C$_{60}$-$^4$He$_N$
clusters are investigated. Particular attention is paid to the growth
of the highly pronounced layered density profile. The evolution
towards bulk liquid and surface thickness at the free interface are
discussed.

FILENAME: fuller.tex

\end{abstract}

\section{Introduction}
\label{sec:introduce}

Since the discovery of the fullerenes and nanotubes a large
amount of work has been devoted to study properties of such
systems.\cite{dress} In particular, in the case of nanotubes the
adsorption of quantum fluids has been thoroughly investigated.
\cite{cab} The physisorption potential for such systems has been
modeled several years ago by Stan and Cole.\cite{stucco}
Calculations performed for $^4$He adsorbed into carbon
nanotubes have shown interesting features.\cite{gate,gate1} It
was found that inside tubes of radius $\lesssim 7 \rm \AA$,
besides the formation of a shell phase located at about 3 $\rm
\AA$ from the wall, a quasi-one-dimensional structure along the
cylinder axis is also developed. These sorts of
quasi-one-dimensional structures are of theoretical interest.
\cite{crow,bon} On the other hand, the shell phase located close
to the wall exhibits a surface density in agreement with the
experimental value\cite{two,brush} obtained for monolayer
completion density on graphite (see Table\ \ref{table1}). Moreover,  
adsorption in bundles of nanotubes have been also studied
showing a rich pattern of phase transitions.\cite{cole,gate2}

On the other hand, hitherto there is no study of the adsorption of
$^4$He onto C$_{60}$ in the literature. One expects that the
cloud of $^4$He surrounding this fullerene will exhibit a shell
structure of the type found for helium clusters doped with atomic
or molecular impurities.\cite{barn,mash,herb,cow,hart,mac,gate3}
It is of interest to determine how pronounced will be the layering
as well as to estimate the shells' surface densities in order to
make a comparison with results for other geometries mentioned
above. By looking at the behavior of these C$_{60}$-$^4$He$_N$
clusters for increasing number of helium atoms, $N$, one may
search whether growth instabilities like those found in the case of
planar graphite substrates\cite{clem} are also present in this case.
Furthermore, one may search for what is going on when the
clusters approach bulk liquid. Other issue to be examined is the
liquid-vacuum interface thickness. There is a current interest in
this observable\cite{sur2} and experimental data for spherical
and planar helium systems are available.\cite{dali,penal}
Moreover, since the field is open for exploration new unexpected
features may appear.
 
The paper is organized in the following way. In Sec.\ \ref{%
sec:model} we outline the model and construct the physisorption
potential for the interaction between C$_{60}$ and $^4$He atoms.
Numerical results are presented in Sec.\ \ref{sec:study} together
with the discussion of the whole picture exhibited by the systems.
Section \ref{sec:conclude} is devoted to final remarks.

\section{The model}
\label{sec:model}

\subsection{Theoretical Framework}
\label{sec:theory}

The adsorption properties at $T=0$ K may be studied by analyzing the
grand thermodynamic potential
\begin{equation}
\Omega = E_{\rm gs} - \mu\,N \:, \label{omega00}
\end{equation}
where $E_{\rm gs}$ is the ground-state energy, $\mu$ the chemical
potential, and $N$ the number of particles of the adsorbed fluid. In the
present work we adopted a density functional (DF) approach, which
has proven to be a successful tool for treating this kind of quantum
many-body problems. In such a theory the ground-state energy of an
interacting $N$-body system of $^4$He atoms, confined by an
adsorbate-substrate potential $U_{\rm sub} ({\bf r})$, may be
written as
\begin{eqnarray}
E_{\rm gs}~&&= -{\hbar^2\over2 m} \int { d{\bf r} \,\sqrt{\rho({\bf r})} \,
{\bf \nabla}^2 \sqrt{\rho({\bf r})}} + \int { d{\bf r}\,\rho({\bf r})\,e_{sc}({\bf r})}
\nonumber\\
&&~~~+ \int { d{\bf r}\,\rho({\bf r})\,U_{\rm sub}({\bf r}) } \;,
\label{Ene0}
\end{eqnarray}
where $\rho({\bf r})$ is the one-body density and $e_{sc}({\bf r})$ the
self-correlation energy per particle. The density profile $\rho({\bf r})$ is
determined from the Euler-Lagrange (EL) equation derived from the
condition
\begin{equation}
\frac{\delta \Omega}{\delta \rho({\bf r})} = \frac{\delta \{\,E_{\rm gs}[\rho,
{\bf \nabla}\rho] - \mu\,N\,\}}{\delta \rho({\bf r})} = 0 \;. \label{vary0}
\end{equation}

In the present work, we adopted for $^4$He the non-local density
functional (NLDF) proposed by the Orsay-Trento (OT) collaboration,
\cite{dale} with only one change which consist in neglecting the
non-local gradient correction to the kinetic energy term as it was done
by Mayol {\it et al.}\cite{may} Although this kind of functionals is known,
we present a summary for the sake of completeness. The
self-correlation energy is expressed as
\begin{eqnarray}
e_{sc}({\bf r}) =~&& \frac{1}{2}\,\int { d{\bf r}\prime \,
\rho({\bf r}\prime)\,V^{\rm OT}_l(\mid {\bf r} - {\bf r}\prime \mid) }
+ \frac{c\prime_4}{2} \, [\,\bar\rho({\bf r})\,]^2 \nonumber\\
&&+ \frac{c\prime\prime_4}{3}  \, [\,\bar\rho({\bf r})\,]^3 \;, \label{dupe2}
\end{eqnarray}
with $c\prime_4=-2.411857 \times 10^4$ K$\rm \AA^6$ and
$c\prime\prime_4=1.858496 \times 10^6$ K$\rm \AA^9$. The
two-body interaction, $V_l^{\rm OT}(\mid {\bf r} - {\bf r}\prime \mid)$,
was taken as the $^4$He-$^4$He Lennard-Jones (LJ) potential
screened in a simple way at distances shorter than a characteristic
distance $h_{\rm OT}$
\begin{eqnarray}
\: V^{\rm OT}_l(r) = \left\{
\begin{array}{lll}
4 \epsilon_{\rm LJ} \biggr[\left(\frac{\sigma_{\rm LJ}}{r}
\right)^{12} - \left(\frac{\sigma_{\rm LJ}}{r}\right)^6\biggr] &
{\rm if} & r \, \geq h_{\rm OT} \;, \\
&& \\
0 & {\rm if} & r \, < h_{\rm OT} \;, \\
\end{array} \right. \label{screen}
\end{eqnarray}
with the standard de Boer and Michels parameters,\cite{boom}
namely, well depth $\epsilon_{\rm LJ} = 10.22$~K and hard core
radius $\sigma_{\rm LJ} = 2.556$~\AA, in addition, the screening
distance $h_{\rm OT} = 2.190\,323$~\AA.

Quantity $\bar{\rho}({\bf r})$ is the ``coarse-grained density'' defined
as the straight average of $\rho({\bf r})$ over a sphere centered 
at ${\bf r}$ and with a radius equal to the screening distance
$h_{\rm OT}$
\begin{equation}
\bar{\rho}({\bf r}) = \int { d{\bf r}\prime \, \rho({\bf r}\prime)
\,\cal{W}(\mid {\bf r} - {\bf r}\prime \mid) } \;, \label{coarse}
\end{equation}
where $\cal{W}(\mid {\bf r} - {\bf r}\prime \mid)$ is the normalized
step function
\begin{eqnarray}
\: \cal{W}(\mid{\bf r} - {\bf r}\prime \mid) = \left\{
\begin{array}{lll}
\frac{3}{4 \pi h_{\rm OT}^3} & {\rm if} & \mid {\bf r} - {\bf
r}\prime \mid \; \leq h_{\rm OT} \;, \\
& & \\
0 & {\rm if} & \mid {\bf r} - {\bf r}\prime \mid \; > h_{\rm OT}
\;. \\
\end{array} \right. \label{the}
\end{eqnarray}

\subsection{The physisorption potential}
\label{sec:potential}

The fullerene C$_{60}$ is to a very good approximation a spherical
system.\cite{dress} For constructing the potential due to this carbon
structure we followed the procedure already applied in the case of
a single nanotube.\cite{stucco} So, we suppose that an adsorbed
$^4$He atom located at $\vec{r}$, measured from the C$_{60}$
center, interacts with a single substrate C atom of the spherical shell
at $\vec{r}\prime$ via an isotropic LJ pair potential with standard
parameters $\varepsilon_{\rm CHe}=16.24$~K and $\sigma_{\rm
CHe} = 2.74$ $\rm \AA$ taken from Ref.\ \onlinecite{brush}. Next, we  
consider a spherically averaged potential by assuming that the C
atoms are uniformly distributed over the spherical surface of radius
$R_{\rm full}$. Under these conditions, the total effect of all carbon
atoms is given by
\begin{eqnarray}
U&&_{\rm sub}(r) = 8\,\pi\,\varepsilon_{\rm CHe}\,\Theta_s\,
R^2_{\rm full} \nonumber\\
&&\times \int^\pi_0 \,\biggr[ \left(\frac{\sigma_{\rm CHe}}{\mid \vec{r}
- \vec{r}\prime \mid} \right)^{12} - \left(\frac{\sigma_{\rm CHe}}{\mid
\vec{r} - \vec{r}\prime \mid} \right)^6 \, \biggr] \sin\theta\,
d\theta \:. \label{sub0}
\end{eqnarray}
Here, $\Theta_s=N_c/(4 \pi R^2_{\rm full})$ is the uniform surface
density of C atoms. After some straightforward algebra the
expression for $U_{\rm sub}(r)$ inside the fullerene can be cast
into the form
\begin{eqnarray}
U_{\rm sub}&&(r) = 2\,N_c\,\varepsilon_{\rm CHe} \nonumber\\
&&\times \biggr[ \left(
\frac{\sigma_{\rm CHe}}{R_{\rm full}} \right)^{12} M_{6}(\nu)
- \left(\frac{\sigma_{\rm CHe}}{R_{\rm full}}\right)^6 M_3(\nu)
\biggr] \:, \label{sub1}
\end{eqnarray}
where $M_n(\nu)$ stands for the integral
\begin{eqnarray}
M_n&&(\nu) = \int^\pi_0 \,\frac{\sin{\varphi}\,d\varphi}
{(1\,+\,\nu^2\,-\,2\,\nu\,\cos{\varphi})^n} \nonumber\\
&&= \int^1_{-1} \,\frac{d x}
{(1\,+\,\nu^2\,+\,2\,\nu\,x)^n}
\nonumber\\
&&= \frac{1}{2(n-1)\nu} \biggr[ \frac{1}{(1-\nu)^{2(n-1)}}
-  \frac{1}{(1+\nu)^{2(n-1)}} \biggr] \;, \label{ems}
\end{eqnarray}
with $\nu=r/R_{\rm full}$. On the other hand, the adsorption
potential outside of the fullerene becomes
\begin{eqnarray}
U_{\rm sub}&&(r) = 2\,N_c\,\varepsilon_{\rm CHe} \nonumber\\
&& \times  \biggr[ \left(
\frac{\sigma_{\rm CHe}}{r} \right)^{12} M_{6}(\nu)
- \left(\frac{\sigma_{\rm CHe}}{r}\right)^6 M_3(\nu) \biggr] \:,
\label{sub2}
\end{eqnarray}
with $\nu=R_{\rm full}/r$. Since in the limit $\nu \to 0$ holds
\begin{equation}
M_n(\nu \to 0) = 2 \;, \label{emus}
\end{equation}
one gets the correct result at the C$_{60}$ center
\begin{equation}
U_{\rm sub}(r=0) = N_c\,4\,\varepsilon_{LJ} \biggr[ \left(
\frac{\sigma_{\rm CHe}}{R_{\rm full}} \right)^{12}
- \left(\frac{\sigma_{\rm CHe}}{R_{\rm full}}\right)^6 \biggr] \:,
\label{center}
\end{equation}
and also the appropriate asymptotic behavior very far away
from the fullerene
\begin{equation}
U_{\rm sub}(r >> R_{\rm full}) = N_c\,4\,\varepsilon_{LJ}
\biggr[ \left(\frac{\sigma_{\rm CHe}}{r} \right)^{12}
- \left(\frac{\sigma_{\rm CHe}}{r}\right)^6 \biggr] \:. \label{far}
\end{equation}
We expect that the resulting potential would give a reliable
description of the main features of the systems.

\section{Results and Analysis}
\label{sec:study}

In the case of a spherical symmetry the variation of Eq.\
(\ref{vary0}) leads to the following Hartree like equation for the
square root of the one-body helium density
\begin{eqnarray}
-\frac{\hbar^2}{2 m} \:&& \left(\,\frac{d^2}{dr^2}
+ \frac{2}{r}\,\frac{d}{dr} \right) \,\sqrt{\rho(r)} \nonumber\\
&&+~\biggr[\,V_H(r) + U_{\rm sub}(r) \biggr] \,\sqrt{\rho(r)}
= \mu \, \,\sqrt{\rho(r)} \;, \label{hairs}
\end{eqnarray}
which also determines $\mu$. Here $V_H({\bf r})$ is a Hartree
mean-field potential given by the first functional derivative of
the total correlation energy $E_{sc}[\rho]$
\begin{equation}
V_H({\bf r}) = \frac{\delta E_{sc}[\rho]}{\delta \rho({\bf r})}
= \frac{\delta}{\delta \rho({\bf r})} \, \int d{\bf r}\prime \,
\rho({\bf r}\prime)\,e_{sc}({\bf r}\prime) \;. \label{harp}
\end{equation}
The expression derived for the spherically symmetric $V_H( r)$ is
given in the Appendix. Equation (\ref{hairs}) was solved at a fixed
number of helium atoms
\begin{equation}
N = 4\,\pi \int^\infty_0 {r^2\;dr\;\rho(r)} \;.
\label{number2}
\end{equation}
The calculations has been carried out for $R_{\rm full} = 3.53$ $\rm
\AA$ (cf. Ref.\ \onlinecite{dress}) that for $N_c=60$ leads to
$\Theta_s=0.38$ ${\rm \AA}^{-2}$ in agreement with the surface
density used for nanotubes. As shown in Fig.\ \ref{fig:potential} the
difference between the inner and outer potential depths is dramatic.

\subsection{Inner Atom}
\label{sec:inside}

Since the radius is very small there is enough room for only one
$^4$He atom inside the fullerene. The solution for such a system is
obtained from Eq. (\ref{hairs}) by using $U_{\rm sub}(r)$ given by
Eq. (\ref{sub1}) and setting to zero the correlation energy. So, the
problem is reduced to a Schr\"odinger equation.

Figure \ref{fig:wave} shows the energies and wave functions for
the lowest two eigenstates together with the inner potential. A
remarkable feature of these results is that inside of the fullerene a
helium atom is extremely strongly bound with a ground-state energy  
$E_{\rm gs} = E_0 = -592$ K.

\subsection{The C$_{60}$-$^4$He$_N$ clusters}
\label{sec:outside}

Adsorption on the external side of the fullerene has been studied by
solving the complete Eq.\ (\ref{hairs}) for systems up to $N=1600$
$^4$He atoms. In this approach we neglect the zero-point motion of  
the fullerene and treat it as a body with infinite mass. It has been
established that this is a reasonable description in the case of helium
clusters doped with sulfur hexafluoride molecules (SF$_6$).\cite{%
hart,mac,gate3} One expects that this approach be even better for
C$_{60}$ because it is a heavier and stronger attractor.

The energy per particle, $e=E_{\rm gs}/N$, and the chemical
potential, $\mu$, are plotted in Fig.\ \ref{fig:energy} as a function of
$N$. The size of these results can be understood by taking into
account that the adsorption potential outside presents a well depth
of about 100 K, being much smaller than inside as shown in Fig. \
\ref{fig:potential}. Examples of density profiles are displayed in Fig.\
\ref{fig:profiles}. Small C$_{60}$-$^4$He$_N$ clusters exhibit a
typical layered structure. Two well defined layers are developed at
$N \simeq 160$. For larger systems begins the transition towards
bulk liquid.

Let us first focus the attention on small systems. After a small
region where $\Omega > 0$ (for $N \lesssim 20$) begins the
growth of a stable monolayer film. For $N \gtrsim 60$ starts the
formation of the second layer and $\mu$ exhibits a kink. At this
point one must be careful because a growth instability similar to
that found for films adsorbed on strongly attractive planar
substrates\cite{clem} might appear. This kind of instability would
cause a jump of $N$ in the adsorption isotherm. We have checked
the possible presence of such a feature by increasing $N$ in steps
of $\Delta N =1$. It was found that $\mu$ is always an increasing
function of $N$ as shown in Fig.\ \ref{fig:growth}. In addition, we
analyzed whether the unstable regime is close. A transition to
instabilities may be induced by increasing the strength of the
C$_{60}$-$^4$He interaction. In turn, this may be achieved by
either changing the potential parameters or going to fullerenes
with more carbon atoms, i.e., larger radii. Since we would like to
stick C$_{60}$, the former way for enhancing the attraction of the
potential was selected. Calculations have been  carried out for
$\sigma_{\rm CHe} = 2.98$ $\rm \AA$, this value has been widely
used in the past for the graphite-helium interaction (see discussion 
in Ref.\ \onlinecite{cole2}) and it increases about 10$\%$ the well
depth. The new plateau corresponding to the growth of the
second layer is shown in Fig.\ \ref{fig:growth}. From this drawing
one concludes that in this case $\mu$ is also an increasing
function of $N$ indicating stability of the solutions. So, we can
state that the solutions for $\sigma_{\rm CHe} = 2.74$ $\rm \AA$
do not undergo growth instability and, in addition, that the unstable
regime is not close to the adopted scenario.

We shall now examine the surface density of the first two layers.
Experimentally it has been determined that for a planar substrate
of graphite the adsorbed superfluid $^4$He presents two well
defined layers with coverages $n_{\rm s1}=0.115$ $\rm \AA^{-2}$
and $n_{\rm s2}=0.093$ $\rm \AA^{-2}$ for the first and second
layer, respectively.\cite{two,brush} In the present work, surface
densities were evaluated according to
\begin{equation}
n_{si} =  \frac{N_i}{4\,\pi\,<R_i>^2} \:, \label{cover}
\end{equation}
where $<R_i>$ is the location of the $i$-layer and $N_i$ its
number of particles. Looking at Fig.\ \ref{fig:profiles} one realizes
that the first layer is well defined, it is centered at $<R_1>=6.41$
$\rm \AA$ and a simple integration yields $N_1=57$ leading to
$n_{\rm s1}=0.110$ $\rm \AA^{-2}$. Although the position of the
second layer may be easily obtained, $<R_2>=9.35$ $\rm  \AA$,
the evaluation of $N_2$ is not straightforward because the
right-hand side region of the peak is not clearly exhibited. By
taking into account the superposition of tails of the second and
third peaks we estimated $N_2=90$ which gives $n_{\rm s2}=
0.082$ $\rm \AA^{-2}$. In order to facilitate the comparison all
the results for the surface density mentioned along the paper are
collected in Table\ \ref{table1}. A glance at this table indicates a
fair agreement between calculated and experimental data,
although the present value of $n_{\rm s2}$ is slightly smaller that
the experimental one, presumably due to the curvature of C$_{60}$. 

For large $N$ the systems evolve towards bulk liquid. This is merely  
an expected behavior. The real theoretical prediction is how rapidly
the clusters approach that limit when $N$ is increased. In addition,
the comparison with other doped clusters becomes of interest. In
Fig.\ \ref{fig:revers} we plotted $e$ and $\mu$ as a function of
$N^{-1/3}$, which is the appropriate expansion parameter for
spherical drops. To fix ideas we recall that $e$ and $\mu$ for free
spherical $^4$He clusters obey the following asymptotic laws:
\begin{equation}
e_{\rm As} = \frac{E_{\rm gs}}{N} = e_B
+ \left(\frac{36\,\pi}{\rho^2_0}\right)^{1/3} \sigma_\infty\,N^{-1/3} \;,
\label{foot}
\end{equation}
and
\begin{equation}
\mu_{\rm As} = \left(\frac{d E_{\rm gs}}{d N}\right)_{\rm As} = e_B
+ 2\,\left(\frac{4\,\pi}{3\,\rho^2_0}\right)^{1/3} \sigma_\infty\,N^{-1/3} \;.
\label{foots}
\end{equation}
Here $e_B=\mu_0=-7.15$ K, $\rho_0=0.021836$ $\rm \AA^{-3}$,
and $\sigma_\infty=0.274$ $\rm K\AA^{-2}$ (see Refs.\ \onlinecite{%
sat2,gest,edda,isis,raw}). These straight lines are included in Fig.\
\ref{fig:revers}. The results displayed in this figure indicate that
$\mu$ reaches much more rapidly the asymptotic trend than $e$.
At about $N \simeq 500$ the chemical potential lies, in practice, on
the straight line given by Eq.\ (\ref{foots}). At this value of $N$ the
slope of $\mu$ changes indicating a phase transition to a
metastable regime. The behavior of $e$ is biased by the very
strongly attractive potential. The dashed curve indicates how $e$
might attain $e_B$. This curve was obtained by fitting data for $N
\geq 500$ to 
\begin{equation}
e = e_B + \left(\frac{36\,\pi}{\rho^2_0}\right)^{1/3} \sigma_\infty\,
N^{-1/3} + a_c\,N^{-2/3}  + a_0\,N^{-1} \;,\label{foul}
\end{equation}
which yielded $a_c=0.2$ K and $a_0=-3316.4$ K. The large
negative value of $a_0$ is caused by the strong attraction
exerted by C$_{60}$. Furthermore, it is important to notice that
these coefficients also explain the behavior of $\mu$. In this
region of $N$ the chemical potential obeys Eq.\ (\ref{foots})
due to the facts that  $a_c$ is small and the huge $a_0$ does
not contribute because $\mu=dE_{\rm gs}/dN$.

The trend of $e$ towards the asymptotic law given by Eq.\ 
(\ref{foot}) is noticeable delayed in comparison with diffusion
Monte Carlo (DMC) results\cite{barn} for SF$_6$-$^4$He$_N$
clusters. It should be mentioned that these DMC values were
reproduced fairly well by NLDF evaluations (see Fig. 1 in Ref.\ 
\onlinecite{gate3}) as well as by hypernetted chain calculations
(cf. Fig. 12 in Ref.\ \onlinecite{cow}). Perhaps it is worthwhile to
notice that the latter figure also indicates that helium doped with
atomic impurities like Ne, Ar, Kr, and Xe reaches even more
rapidly the asymptotic behavior for $e$ than the
SF$_6$-$^4$He$_N$ clusters do. On the other hand, from a
glance at Fig. 3 of Ref.\ \onlinecite{gate3} one could conjecture
that the collective energy of the dipole mode for
C$_{60}$-$^4$He$_N$ clusters with $N$ even a few hundreds
bigger than $1000$ would still not approach zero. This means
that the instability scenario for the dopant would not be reached
for $N=1600$ yet.

Let us briefly refer to the entrance to the bulk regime.
From the fit to Eq.\ (\ref{foul}) it is possible to determine that the
crossing between $e$ and $\mu$ takes place at $N^{-1/3}
\simeq 0.041$, see also Fig.\ \ref{fig:revers}. This indicates that
at $N \simeq 14500$ the bulk could be reached. It would be
interesting to find out whether the collective energy of the
dipole mode for such large systems is still positive.

Finally, we shall focus our attention on the width of the
liquid-vacuum interface. At this interface, the density profile of
$^4$He changes continuously from liquid density $\rho_\ell$ to
zero over a distance of some {\aa}ngstr\"oms. The width $W$ of
a surface is defined as the distance in which the density
decreases from $0.9\rho_\ell$ to $0.1\rho_\ell$. For consistency
with other studies we adopted $\rho_\ell=\rho_0$. The analysis
was centered on systems with $N \geq 600$ because for such
clusters the oscillations of $\rho(r)$ close to the free interface are
already damped enough. The present results for $W$ are
compared in Fig.\ \ref{fig:width} with values calculated for free
$^4$He droplets\cite{sur2} and experimental data.\cite{dali,%
penal} The results obtained in the measurement of large droplets
published in Ref.\ \onlinecite{dali} are represented by the reported
mean thickness $W = 6.4 \pm 1.3$~\AA. In the case of Ref.\
\onlinecite{penal} only the two data included by the authors in the
abstract are shown. Since the latter values correspond to rather
broad planar films, we plotted them schematically close to origin
for the abscissa. The overall agreement exhibited in Fig.\ 
\ref{fig:width} is good.

\section{Final remarks}
\label{sec:conclude}

The main results of this investigation about C$_{60}$-$^4$He$_N$
clusters may be summarized in the following way. Only one $^4$He
atom may be put inside of this fullerene and its binding energy is
very large. The energetics of the adsorption on the external side
was studied and the structure of the films was determined. 

The evolution of the systems for increasing $N$ does not exhibit
any growth instability after the completion of the first layer and, in
addition, the unstable scenario is not close. The surface density of
the first two layers are in fair agreement with experimental data for
planar graphite. The analysis of the transition towards bulk liquid
as a function of $N$ shows that $\mu$ attains more rapidly the
asymptotic behavior than $e$. Results obtained from a fit of $e$
account for this feature. The C$_{60}$-$^4$He$_N$ clusters
approach the bulk regime much slower than helium clusters doped
with a SF$_6$ molecule.

The surface thickness at the liquid-vacuum interface obtained
for rather large clusters matches reasonably well with theoretical
evaluations for free helium spheres and experimental data
for droplets and films.

For the future work it remains the study of excitations, which may
give information on the location of C$_{60}$. Furthermore, this
issue is also a challenge for experimental work as well as Monte
Carlo simulations due to the fact that such investigations may
also shed some light on the importance of the zero-point motion
of this heavy dopant. Moreover, it would be exciting to look for a
possible levitation of C$_{60}$ in bulk helium.

\acknowledgements

The authors acknowledge enlightening discussions with E.S.
Hern\'andez and thank H. Bonadeo and E. Burgos for providing
valuable information.
This work was supported in part by the Ministry of Culture and
Education of Argentina through Grants PICT-2000-03-08450 from
ANPCYT and No. X103 from University of Buenos Aires.

\appendix
\section*{Hartree Mean-Field Potential}

In this Appendix we compile the relevant expressions derived in the
adopted NLDF approach. For a spherical geometry the definition of
the Hartree mean-field of Eq.\ (\ref{harp}) leads to
\begin{eqnarray}
V_H&&(r) = \int d{\bf r}\prime \, \rho(r\prime) \, V^{\rm OT}_l
(\mid{\bf r} - {\bf r}\prime \mid)) \nonumber\\
&&+ \,\frac{c\prime_4}{2}\,[\,\bar{\rho}(r)\,]^2
+ \,\frac{c\prime\prime_4}{3}\,[\,\bar{\rho}(r)\,]^3 \nonumber\\
&&+ \int d{\bf r}\prime
\rho(r\prime) \{ c\prime_4 \,\bar{\rho}(r\prime)\,+ c\prime\prime_4 \,
[\,\bar{\rho}(r\prime)\,]^2\,\}\,\cal{W}(\mid {\bf r} - {\bf r}\prime \mid) \;.
\nonumber\\  \label{harp0}
\end{eqnarray}
Let us first provide the expressions of the contributions
involving the ``coarse-grained density'' $\bar{\rho}(r)$, {\it i.e.}, 
\begin{equation}
\bar\rho(r) = \int d{\bf r}\prime \, \rho(r\prime) \,
\cal{W}(\,\mid{\bf r} - {\bf r}\prime \mid) \;, \label{rbar}
\end{equation}
and
\begin{equation}
\bar\rho_V(r) = \int d{\bf r}\prime \, \rho(r\prime) \,
[\,\bar{\rho}(r\prime)\,]^n \, \cal{W}(\,\mid{\bf r} -
{\bf r}\prime \mid) \;. \label{Vrbar}
\end{equation}
Both these integrals may be cast into the  form
\begin{equation}
\bar{\cal{R}}(r) = \frac{3}{4 \pi h^3_{\rm OT}} \int d{\bf r}\prime
\, {\cal{R}}(r\prime) \, \Theta(\,h_{\rm OT}\,- \mid{\bf r} - {\bf
r}\prime \mid) \;. \label{rrbar}
\end{equation}
After introducing spherical coordinates and taking into account
that the step function is symmetric in the azimuthal angle
$\varphi$, the integration over this variable yields
\begin{eqnarray}
\bar{\cal{R}}(r)~&&= \frac{3}{2\,h^3_{\rm OT}}
\int^{r\prime_{\rm max}}_{r\prime_{\rm min}} r\prime^2\,dr\prime \,
{\cal{R}}(r\prime)\,\int^{\theta_{\rm max}}_0 \sin\theta \, d\theta
\nonumber\\
&&~~~\times \Theta[\,h^2_{\rm OT} - r^2 - r\prime^2
+ 2\,r\,r\prime\,\cos\theta\,] \;. \label{rbar0}
\end{eqnarray}
Since we shall treat helium systems located at $r > h_{\rm OT}$,
the limits for the radial integral become $r\prime_{\rm min} = r -
h_{\rm OT}$ and $r\prime_{\rm max}=r+h_{\rm OT}$.
Furthermore, the step function sets
\begin{equation}
\cos \theta_{\rm max} = \frac{r^2 + r\prime^2 - h^2_{\rm OT}}{2
\,r\,r\prime} \;. \label{cos}
\end{equation}
Hence, one gets\cite{sur2}
\begin{equation}
\bar{\cal{R}}(r)
= \frac{3}{4\,r\,h_{\rm OT}} \int^{r+h_{\rm OT}}_{r-h_{\rm OT}}
r\prime\,dr\prime\,{\cal{R}}(r\prime)\,\biggr[\,1
- \left(\frac{r-r\prime}{h_{\rm OT}}\right)^2 \, \biggr] \;.
\label{rbar3}
\end{equation}
On the other hand, the integration of the screened LJ potential
contributing to Eq.\ (\ref{harp0}) for $r \geq h_{\rm OT}$ leads to
\begin{eqnarray}
V&&^{\rm LJScr}_H(r) = \int d{\bf r}\prime \, \rho(r\prime) \,
V^{\rm OT}_l (\mid{\bf r} - {\bf r}\prime \mid)) \nonumber\\
&&= \frac{4 \pi \epsilon_{\rm LJ}\,\sigma^2_{\rm LJ}}{r} \biggr\{
\int^{r-h_{\rm OT}}_0 + \int^\infty_{r+h_{\rm OT}} \biggr\}
r\prime\,dr\prime \, \rho(r\prime) \nonumber\\
&&~~~~~~~~~ \times \biggr\{ \frac{1}{5}
\biggr[ \left(\frac{\sigma_{\rm LJ}}{r-r\prime}\right)^{10}
- \left(\frac{\sigma_{\rm LJ}}{r+r\prime}\right)^{10} \biggr]
\nonumber\\
&&~~~~~~~~~~ - \frac{1}{2} \biggr[ \left(\frac{\sigma_{\rm LJ}}
{r-r\prime}\right)^4 - \left(\frac{\sigma_{\rm LJ}}{r+r\prime}
\right)^4 \biggr] \biggr\} 
\nonumber\\
&&~~+ \,\frac{4 \pi \epsilon_{\rm LJ}\,\sigma^2_{\rm LJ}}{r}
\,\int^{r+h_{\rm OT}}_{r-h_{\rm OT}} r\prime\,dr\prime\,
\rho(r\prime)\,\biggr\{ \frac{1}{5} 
\biggr[ \left(\frac{\sigma_{\rm LJ}}{h_{\rm OT}}\right)^{10}
\nonumber\\
&&~~~~~~~~ - \left(\frac{\sigma_{\rm LJ}}{r+r\prime}
\right)^{10} \biggr] - \frac{1}{2} \biggr[ \left(\frac{\sigma_{\rm LJ}}
{h_{\rm OT}}\right)^4 - \left(\frac{\sigma_{\rm LJ}}{r+r\prime}
\right)^4 \biggr]\biggr\} \;. \nonumber\\
\label{VHar3}
\end{eqnarray}

It is worthwhile to notice that we expressed these quantities in a
rather simple compact form similar to Eq.\ (2.16) written in Ref.\
\onlinecite{dhpt} for planar films instead of adopting the expansion
in terms of Legendre polynomials proposed in Ref.\
\onlinecite{mash}. The latter procedure is more appropriate for
studying excitations of a given system.

\newpage

\begin{table}
\caption{Surface density of $^4$He layers adsorbed on planar
graphite compared with results for helium shells confined into a
cylindrical carbon nanotube and adsorbed onto a spherical
C$_{60}$.}
\begin{tabular}{lcccr}
Geometry & Method & $n_{\rm s1}$~[\AA$^{-2}$]
& $n_{\rm s2}$~[\AA$^{-2}$] & Ref. \\
\tableline
Planar & Experiment & 0.115 & 0.093 & \onlinecite{two,brush} \\
Cylindrical & Theory  & 0.113 &            & \onlinecite{gate} \\
Spherical   & Theory & 0.110  & 0.082 & PW$^a$ \\
\end{tabular}
$^a$ PW stands for present work.
\label{table1}
\end{table}

\newpage

\begin{figure}
\caption{The left-most curve is the potential for a $^4$He atom
inside the fullerene evaluated with Eq.\ (\protect\ref{sub1}). The
right-most curve is the potential for a $^4$He atom outside of the
fullerene evaluated with Eq.\ (\protect\ref{sub2}). Notice that the
surface of C$_{60}$ is located at $r=3.53$ $\rm \AA$.}
\label{fig:potential}
\end{figure}

\begin{figure}
\caption{Energies and wave functions of the two lowest eigenstates
for a single $^4$He atom inside C$_{60}$. Quantities $\psi_n(r)$
are in arbitrary scale. The bold curve is the potential.}
\label{fig:wave}
\end{figure}

\begin{figure}
\caption{Energy per particle and chemical potential for
C$_{60}$-$^4$He$_N$ clusters as a function of the number of
helium atoms.}
\label{fig:energy}
\end{figure} 

\begin{figure}
\caption{Density profiles for a series of C$_{60}$-$^4$He$_N$
clusters. Results for $N=50,$ 100, 150, 200, 300, 400, ..., up to
1600 are displayed.}
\label{fig:profiles}
\end{figure} 

\begin{figure}
\caption{Chemical potential as a function of $N$. The plateau
corresponding to the growth of the second layer is shown for two
values of $\sigma_{\rm CHe}$ (see text).}
\label{fig:growth}
\end{figure} 

\begin{figure}
\caption{Energy per particle (filled circles) and chemical potential
(empty circles) as a function of $N^{-1/3}$. Asymptotic laws for
$e$ and $\mu$ are displayed. The dashed curve indicates the
trend of $e$ towards the asymptotic $e_B=\mu_0$. Stars are
DMC values for the energy per particle of SF$_6$-$^4$He$_N$
clusters taken from Ref.\ \protect\onlinecite{barn}, the dot-dashed
curve stands only to guide the eye.}
\label{fig:revers}
\end{figure} 

\begin{figure}
\caption{Thickness of the free surface as a function of $N^{-1/3}$.
Triangles are data for C$_{60}$-$^4$He$_N$ clusters with $N
\geq 600$. Full circles are results for free $^4$He droplets from
Ref.\ \protect\onlinecite{sur2} and the solid curve indicates the
trend of these data. The ``otimes'' ($\otimes$) and the ``oplus''
($\oplus$) are experimental values from Refs.\
\protect\onlinecite{dali} and \protect\onlinecite{penal},
respectively.}
\label{fig:width}
\end{figure}

\end{document}